\documentclass[12pt,a4paper]{article}
\usepackage{amsfonts,latexsym}
\usepackage{amsmath,amssymb}
\usepackage{graphicx,color}

\renewcommand{\thefootnote}{\fnsymbol{footnote}}

\begin{document}

\vspace{12mm}

\begin{center}
{{{\Large {\bf Extended thermodynamic analysis  of a charged Horndeski black hole }}}}\\[10mm]

{Yun Soo Myung\footnote{e-mail address: ysmyung@inje.ac.kr}}\\[8mm]

{Center for Quantum Spacetime, Sogang University, Seoul 04107, Republic of  Korea\\[0pt] }

\end{center}
\vspace{2mm}

\begin{abstract}
We perform the thermodynamic analysis of a charged Horndeski black hole (CHB) with mass $m$ and charge $q$ obtained from the Einstein-Horndeski-Maxwell theory.
There are two solution branches: one is for the CHB and the other is for the naked singularity (NS).
Thermodynamic behavior for the CHB is similar to that for the Reissner-Nordstr\"{o}m black hole but its Helmholtz free energy is always positive.
If the NS point is included as an extremal point, then the Helmholtz free energy is always negative, implying that the globally stable region is achieved anywhere.
For the NS, its temperature has a maximum point, its heat capacity remains negative without having Davies point, and its free energy decreases without limitation as the charge  $q$ increases.
\end{abstract}
\vspace{5mm}

\vspace{1.5cm}

\hspace{11.5cm}{Typeset Using \LaTeX}
\newpage
\renewcommand{\thefootnote}{\arabic{footnote}}
\setcounter{footnote}{0}

%%%% Introduction %%%%

\section{Introduction}
Horndeski gravity~\cite{Horndeski:1974wa} was  considered  as the most general scalar-tensor gravity of avoiding Ostrogradsky instability.
Among  many kinds of Horndeski gravity, the significant one  is the nonminimal derivative coupling between scalar and Einstein tensor.
Several black hole solutions  were obtained from this gravity~\cite{Rinaldi:2012vy,Anabalon:2013oea,Maselli:2015yva,Babichev:2016fbg,Babichev:2017guv} including a charged Horndenski black hole (CHB) with scalar hair~\cite{Cisterna:2014nua}.

It is known that thermodynamic description of black holes is universal and can be  applicable to black holes found in beyond Einstein gravity.
However, it was found that the Wald's entropy formula and /or the first law of thermodynamics does not  work for the CHB~\cite{Feng:2015wvb}.
To achieve the first law with the ordinary Hawking temperature, they introduced a new term $\Phi_\phi$ of the chemical potential  and a conserved scalar charge $Q_\phi$.
This derivation of the first law  was based on the Wald's entropy ($S_W=\pi(r_+^2-q^2/2)$)  from the Noether's theorem~\cite{Wald:1993nt,Iyer:1994ys}  and the Hawking temperature ($T_\kappa=\kappa/2\pi$) obtained from the surface gravity.
However, Noether's theorem and Wald formula have ambiguities for the Einstein-Horndeski-Maxwell theory.  These  were  fixed by deriving the new temperature and introducing the Bekenstein-Hawking entropy~\cite{Hajian:2020dcq}, leading to the standard first law of thermodynamics. Also, this case satisfies the Smarr formula of its integral form.
We stress that this happened to the Bardeen regular black hole without scalar hair obtained  from the Einstein-nonlinear electromagnetic theory~\cite{Ma:2014qma,Zhang:2016ilt}.
Recently, the relevant astrophysical and physical  implications of the  CHB were discussed in~\cite{Wang:2019cuf,Gao:2023mjb,Li:2024xyu,Liang:2024ygf}.
The thermodynamic aspects of a charged Hordenski black hole were also studied in~\cite{Stetsko:2018fzt,Stetsko:2023jtq}.

In this work, we wish to  perform the extended thermodynamic analysis of a CBH with mass $m$ and charge $q$ obtained from the Einstein-Horndeski-Maxwell theory.
The presence of the secondary scalar hair $\phi(r)$ shows  the overcharge $q\in[0,1.06]$   and  the existence of the naked singularity (NS).
Importantly, this CHB with scalar hair proves the no scalar-haired Cauchy (inner) horizon theorem because the inner horizon disappeared~\cite{Devecioglu:2021xug}.
We include two solution branches: one is for the CHB and the other is for the NS  because  the NS emerges from this theory naturally.
We compute all thermodynamic quantities for the CHB and NS to prove the first law of thermodynamics and the Smarr formula.

For comparison, it is desirable to involve thermodynamics of the Reissner-Nordstr\"{o}m black hole (RNB) obtained from the Einstein-Maxwell theory but the RNB is disconnected to the CHB and the NS. Then, we have three branches for extended thermodynamic analysis. The RNB is considered as a standard form for charged black holes.
Thermodynamic behavior for the CHB without the NS point is similar to that for the RNB, but its Helmholtz free energy is always positive.
If the NS  point is included as an extremal point, then the Helmholtz free energy is always negative, suggesting that the globally stable region is achieved anywhere.
For the NS thermodyamics, its positive temperature has a maximum point and  its heat capacity remains negative without having Davies point.

\section{CHB and NS from the EHM theory}

Firstly, we  introduce  the Einstein-Horndeski-Maxwell theory with a coupling constant $\gamma$~\cite{Cisterna:2014nua,Feng:2015wvb,Hajian:2020dcq}
\begin{equation}\label{Horndeski-L}
\mathcal{L}_{\rm EHM}=\frac{1}{16\pi G}\Big(R-F_{\mu\nu}F^{\mu\nu}+2\gamma G^{\mu\nu}\partial_\mu\phi\partial_\nu\phi\Big)
\end{equation}
Here, $G_{\mu\nu}$ denotes the Einstein tensor and  $F_{\mu\nu}=\partial_\mu A_\nu-\partial_\nu A_\mu$ is the electromagnetic field strength.
This theory shows  an asymptotically flat charged Horndeski black hole (CHB) solution,
\begin{equation}\label{metric-ansatz}
ds^2_{\rm CBH}=-h(r)dt^2+\frac{dr^2}{f(r)}+r^2(d\theta^2+\sin^2\theta d\varphi^2),
\end{equation}
where two metric functions $h(r)$ and $f(r)$ are given by
\begin{align}
h(r)=1-\frac{2m}{r}+\frac{q^2}{r^2}-\frac{q^4}{12r^4}\equiv 1-\frac{2m_f(r)}{r}, \quad f(r)=\frac{4r^4h(r)}{(2r^2-q^2)^2}
\end{align}
with the mass function
\begin{equation}
m_f(r)=m-\frac{q^2}{2r}+\frac{q^4}{24r^3}.
\end{equation}
This mass function is positive for $0<q<\frac{3m}{2}$, zero for $q=\frac{3m}{2}$, and negative for $q>\frac{3m}{2}$.

From $h(r)=0$, one finds four real roots  as
\begin{eqnarray}
&&r_{+1}(m,q)=  \frac{1}{2}\Bigg(m+\frac{\xi}{\sqrt{3}}+\sqrt{\chi+\frac{8\sqrt{3}m(m^2-q^2)}{4\xi}}\Bigg),    \label{h-11} \\
&&r_{+2}(m,q)=  \frac{1}{2}\Bigg(m-\frac{\xi}{\sqrt{3}}+\sqrt{\chi-\frac{8\sqrt{3}m(m^2-q^2)}{4\xi}}\Bigg),   \label{h-22}\\
&&r_{+3}(m,q)=  \frac{1}{2}\Bigg(m-\frac{\xi}{\sqrt{3}}-\sqrt{\chi-\frac{8\sqrt{3}m(m^2-q^2)}{4\xi}}\Bigg),   \label{h-33}\\
&&r_{+4}(m,q)=  \frac{1}{2}\Bigg(m+\frac{\xi}{\sqrt{3}}-\sqrt{\chi+\frac{8\sqrt{3}m(m^2-q^2)}{4\xi}}\Bigg)   \label{h-44}
\end{eqnarray}
with
\begin{equation}
\xi=\sqrt{3m^2-2q^2+q^{4/3}(-9m^2+8q^2)^{1/3}},\quad \chi=2m^2-\frac{4q^2}{3}-\frac{1}{3}q^{4/3}(-9m^2+8q^2)^{1/3}.
\end{equation}
For $m=1$ case, $r_{+1}(1,q)$ and $r_{+2}(1,q)$  describe the  CHB and  the NS, whereas $r_{+3}(1,q)$ is a negative line and $r_{4+}(1,q)$ represents a point at $q=q_{\rm NS}=3/2\sqrt{2}$.
Hence, the former two (CHB and the NS) will be used to compute thermodynamic quantities and prove the first law of thermodynamics and Smarr formula.
\begin{figure*}[t!]
   \centering
   \includegraphics[width=0.5\textwidth]{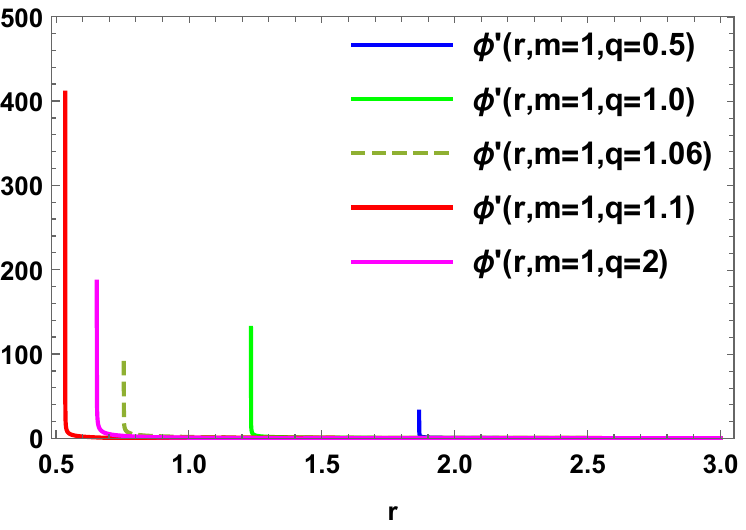}
\caption{Derivative of scalar hair $\phi'(r,m=1,q)$ with $q=0.5,1.0,1.06,1.1,2$  as a function of $r\in[r_+,3]$. They are finite  at the event horizon at $r= r_{+1}(=1.87,1.23,0.76)$ and grow at $r=r_{+2}(=0.54,0.65)$.   }
\end{figure*}
The gauge field and scalar hair take the forms
\begin{align}
A= \left(\frac{q}{r}-\frac{q^3}{6r^3}\right)dt, \quad \phi'(r)=\sqrt{\frac{-q^2}{2\gamma r^2 f(r)}}\to \phi(r)=\frac{q}{r}\int \frac{dr}{\sqrt{-2\gamma f(r)}}.
\end{align}
A real $\phi$ is guaranteed  for $\gamma<0$ and it is secondary but it is not primary because it does not contain any independent scalar charge~\cite{Herdeiro:2015waa}.
 Here, we takes  $\gamma=-2$ for simplicity. As is shown in Fig. 1, its derivative is  regular at the event horizon $r=r_{+1}$ but it seems to grow at $r=r_{+2}$. However, we fail to obtain the real scalar hair from   $\phi(r)$ for $\gamma=-2$.

We observe two singularities from $h(r)$ and $f(r)$: one is at $r=0$ and the other is at $r=r_{\rm NS}(q)=q/\sqrt{2}$, leading to the divergence of Kretschmann scalar defined by $R_{\mu\nu\rho\sigma}R^{\mu\nu\rho\sigma}$.
The weak cosmic censorship conjecture which states that the naked singularity (NS) is behind the horizon implies the condition for mass and charge
\begin{equation}
0<\frac{q}{m}<\frac{q_{\rm NS}}{m}=\frac{3}{2\sqrt{2}m}=\frac{1.06066(\simeq1.06)}{m}.
\end{equation}
 For $q<q_{\rm NS}$, one finds the CHB, while for $q>q_{\rm N}$, one has the NS. The $q=q_{\rm NS}$ corresponds to the singular and extremal point because of $r_{1+}(1,q_{\rm NS})=r_{2+}(1,q_{\rm NS})=r_{4+}(1,q_{\rm NS})$.
\begin{figure*}[t!]
   \centering
   \includegraphics[width=0.5\textwidth]{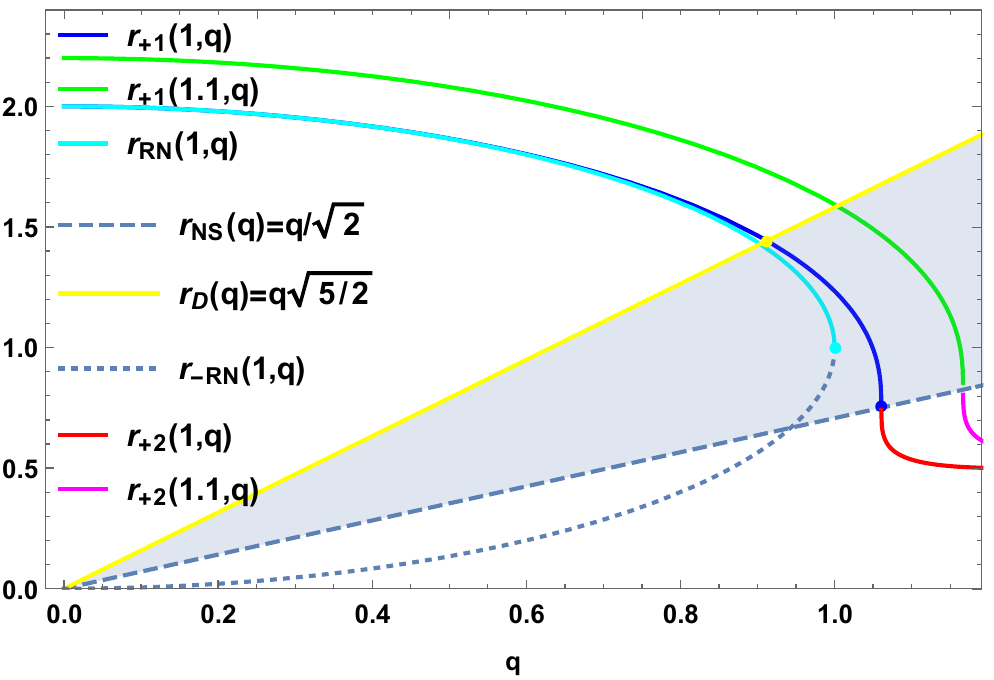}
\caption{ Two horizons (branches) $r_{+1}(m=1,q)$ and $ r_{+2}(1,q)$ and two horizons $r_{\rm RN}(1,q)$ and $r_{\rm -RN}(1,q)$ with an extremal point (cyan dot) at ($q=1,r_{\rm eRN}=1$)  for RNB. Here, we introduce  a line of $r_+= r_{\rm NS}(q)$ for showing  the NS  and $r_+=r_{ D}(q)$ for representing  Davies line of heat capacity. For $r_+> r_{\rm NS}(q)$, one finds the CHB, whereas one has the NS for $r_+<r_{\rm NS}(q)$. One finds the NS point (blue dot) at ($q=1.06,r_{+4}=3/4$).
 The shaded region represents the positive heat capacity for the CHB, while $r_{+}>r_{\rm D}(q)$ denotes the negative heat capacity for the CHB. Finally, we display $r_{+1}(m=1.1,q)$ and $r_{+2}(m=1.1,q)$.  }
\end{figure*}
For a reference, it would be better to introduce the  RNB obtained from the Einstein-Maxwell theory  whose two (outer/inner) horizons and electric potential at the horizon are given by
\begin{equation}
r_{\rm RN}=m+\sqrt{m^2-q^2},\quad r_{\rm -RN}=m-\sqrt{m^2-q^2},\quad \Phi_{\rm RN}=\frac{q}{r_{\rm RN }},
\end{equation}
which implies  a condition for the existence of  two horizons: $0<q/m<1$.
\begin{table}[h]
\begin{tabular}{|c|c|c|c|c|c|c|c|c|c|c|c|}
  \hline
  $q$& 0&0.2  &0.3&0.4&0.5&0.6&0.7&0.8&0.9& 1.0 &1.06\\ \hline
  $r_{+1}(1,q)$ & 2&1.97981&1.954&1.917&1.867&1.802&1.719&1.61&1.46&1.23&0.84 \\ \hline
  $r_{\rm RN}(1,q)$&2&1.9798&1.9539&1.916&1.866&1.8&1.714&1.6&1.44&1.0&N.A. \\ \hline
\end{tabular}
\caption{ Comparing $r_{+1}(1,q)$ with $r_{\rm RN}(1,q)$.  }
\end{table}

Observing  Fig. 2, there is no inner horizon which states that it satisfies no scalar-haired inner horizon theorem~\cite{Devecioglu:2021xug}.
 We find from Table 1 that  $ r_{+1}(1,q\in[0,1.06])\ge r_{\rm RN}(1,q\in[0,1]) $, showing no overlapping  except for the $q=0$.
Its $q$-range is beyond $[0,1]$ of RNB (overcharged), which suggests that the CHB has a scalar hair $\phi(r)$~\cite{Herdeiro:2018wub,Myung:2018vug}.
Instead, it is important to note that  there exists the other branch described by $r_{+2}(m,q\in[1.07,2])$. However, it is always  less than $r_{\rm NS}(q)=q/\sqrt{2}$ for each $q$ and thus, it describes the NS.
The blue dot denotes a point of $(q_{\rm NS}=1.06,r_{+1}(q_{\rm NS})=0.75)$ for $m=1$ which corresponds to  the singular point as well as the extremal point if one takes into account  the other branch $r_{+2}(m,q)$ seriously.
In other words, $r_{+1}(1,q)$ meets $r_{+2}(1,q)$ at the blue dot, while $r_{\rm RN}(1,q)$ meets $r_{\rm -RN}(1,q)$ at the cyan dot, leading to that two points are different.
In principle, this blue point will be excluded,  which is  distinguished  from the extremal RNB  for thermodynamic analysis of the RNB.
In this work, we wish to consider  this blue point seriously  for  studying the Helmholtz free energy.  Finally, a yellow dot at $q=q_{\rm D}=0.91$ represents the Davies point where heat capacity blows up.
For $m\not=1$, we display $r_{+1}(m=1.1,q)$ and $r_{+2}(m=1.1,q)$, showing qualitatively similar behaviors as   $r_{+1}(1,q)$
 and $r_{+2}(1,q)$.

\section{Extended thermodynamic analysis}
The extended thermodynamic analysis means that we  consider the CHB as well as the  NS in the thermodynamic aspect by including two horizons $r_{+1/2}(m,q)$.
Computing  conserved charges yields the mass and electric charge $q$  with $G=1$
\begin{equation} \label{mass-f}
M_1(m,q)=\frac{r_{+1}(m,q)}{2}\Big(1+ \frac{q^2}{r_{+1}^2(m,q)}-\frac{q^4}{r_{+1}^4(m,q)}\Big).
\end{equation}
Also, the temperature from the surface gravity and  electric potential at the horizon  are given by \cite{Feng:2015wvb},
\begin{equation}
T_{\kappa}(m,q)=\frac{\kappa}{2\pi}=\frac{2r^2_{+1}(m,q)-q^2}{8\pi r^3_{+1}(m,q)}, \quad \Phi_{+1}(m,q)=\frac{q}{r_{+1}(m,q)}-\frac{q^3}{6r^3_{+1}(m,q)}.
\end{equation}
We note that  the temperature $T_{\kappa}(m,q)$ together the Wald's entropy [$S_1(m,q)=\pi r_{+1}^2(m,q)$: area law entropy after considering a non-vanishing ambiguity $\pi q^2/2$ with $\gamma=-2$] does not lead  the first law of thermodynamics as $dM=T_{\kappa}dS_1+\Phi_{+1}dq$.
Therefore, we have to  introduce the new temperature by considering ${\cal G}=1+2\gamma X$ with $X=-(\partial\phi)^2/2$ as~\cite{Hajian:2020dcq}
\begin{equation}\label{ex1 T}
T_{1}(m,q)= ({\cal G}-2X{\cal G}')\Big|_{r=r_{+1}}T_{\kappa}= \Big(1-\frac{q^2}{2r^2_{+1}(m,q)}\Big)T_{\kappa}=\frac{(2r_{+1}^2(m,q)-q^2)^2}{16\pi r_{+1}^5(m,q)}.
\end{equation}
\begin{figure*}[t!]
   \centering
  \includegraphics[width=0.4\textwidth]{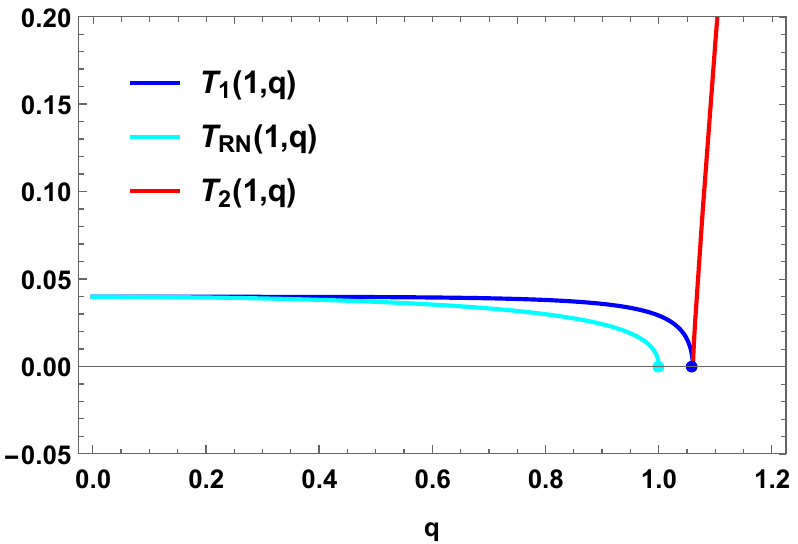}
 \hfill%
    \includegraphics[width=0.4\textwidth]{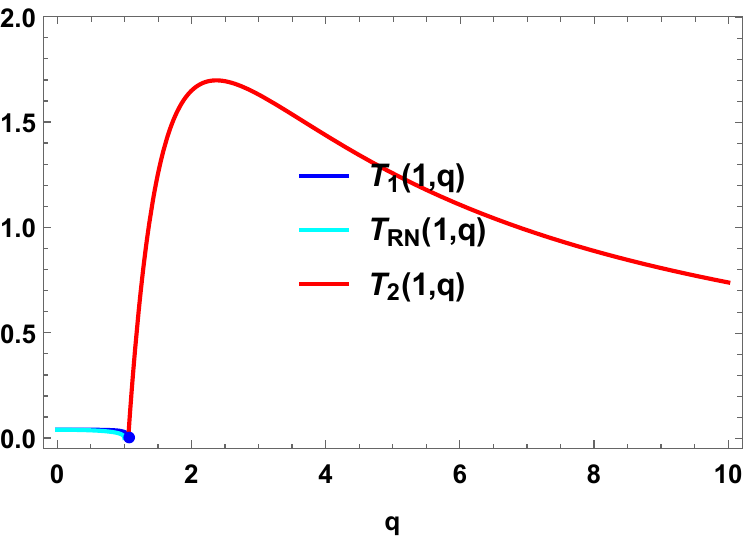}
\caption{(Left) Three  temperatures $T_1(1,q),T_2(1,q),T_{\rm RN}(1,q)$ with $m=1$ as functions of $q\in[0,1.2]$. These are for the CHB. (Right) Temperature $T_2(1,q)$ with $m=1$ as a function of $q\in[0,10]$ for the NS. }
\end{figure*}
It is important to note that this temperature could be obtained from $\partial M(S_1,q)/\partial S_1|_q$ after inserting $r_{1+}(m,q)$ by $\sqrt{S_1/\pi}$ in Eq.(\ref{mass-f}).
With the area entropy   and temperature $T_{1}$, one  verifies  that the first law
\begin{equation}
dM(m,q)=T_{1}(m,q) dS_1(m,q)+\Phi_{+1}(m,q)dq
\end{equation}
is satisfied. Also, it is clear that the Smarr formula expressing the relation between the thermodynamic quantities~\cite{Hajian:2021hje}
 is satisfied  as
\begin{equation}
M(m,q)=2T_{1}(m,q)S_1(m,q)+\Phi_{+1}(m,q) q.
\end{equation}
\begin{figure*}[t!]
   \centering
  \includegraphics[width=0.5\textwidth]{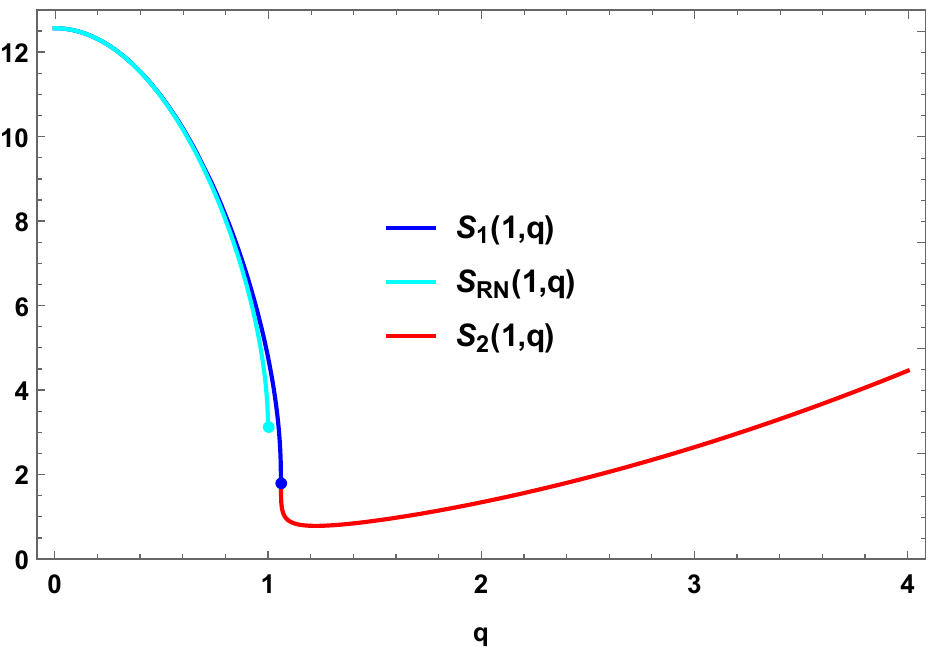}
\caption{ Three  entropies $S_1(1,q),S_2(1,q),S_{\rm RN}(1,q)$ with $m=1$ as functions of $q\in[0,4]$. $S_1(1,q)$ is  for the CHB and $S_2(1,q)$ is for the NS. Here, one finds an inequality of $S_{\rm RN}(1,q)\le S_1(1,q)$ for $q\in[0,1]$. }
\end{figure*}
To achieve  the extended thermodynamics, we  need to define the mass, area entropy, and  new  temperature for  $r_{+2}(m,q)$ so that  its first law and Smarr formula are satisfied.
The relevant quantities for RNB are given as with the outer horizon $r_{\rm RN}(m,q)=m+\sqrt{m^2-q^2}$
\begin{eqnarray}
M_{\rm RN}(m,q)&=&\frac{r_{\rm RN}(m,q)}{2}\Big(1+\frac{q^2}{r_{\rm RN}^2(m,q)}\Big),\quad S_{\rm RN}(m,q)=\pi r_{\rm RN}^2(m,q), \label{tempRN}\\
 T_{\rm RN}(m,q)&=&\frac{1}{4\pi}\Big(\frac{1}{r_{\rm RN}(m,q)}-\frac{q^2}{r_{\rm RN}^3(m,q)}\Big). \nonumber
\end{eqnarray}
It is worth noting that these satisfy the first law and Smarr formula.
 As is shown in Fig. 3, the CHB temperature starts with $T_1(1,0)=1/8\pi$ (Schwarzschild temperature) and  it shows a similar behavior for $q\in[0,1.06]$ as a decreasing function of the RNB.  On the other hand, the positive NS temperature has a maximum point $(q=2.37,T_2(1,q)=1.70)$ and approaches  zero as $q$ increases. This is hotter than that of the CHB. We note that the CHB and RNB extremal temperatures (blue and red dots) show zero temperatures.

Fig. 4 of entropy graphs shows that there exist three branches: the CHB is similar to the RNB but  $S_{\rm RN}(1,q)\le S_1(1,q)$, while the NS shows a quite different behavior of an increasing function of $q$.
The  blue and cyan points represent  different entropies for the extremal CHB and RNB, respectively.
This picture is similar to the entropy picture for scalarized charged black holes obtained from the Einstein-Maxwell-scalar theory with a nonminimal coupling function $f(\Phi)=1+\alpha \Phi^4$, including the hot, the cold, and the RNB branches~\cite{Blazquez-Salcedo:2020nhs}. A difference is that the scalarized charged black holes do not contain  the  NS point and NS.
\begin{figure*}[t!]
   \centering
  \includegraphics[width=0.4\textwidth]{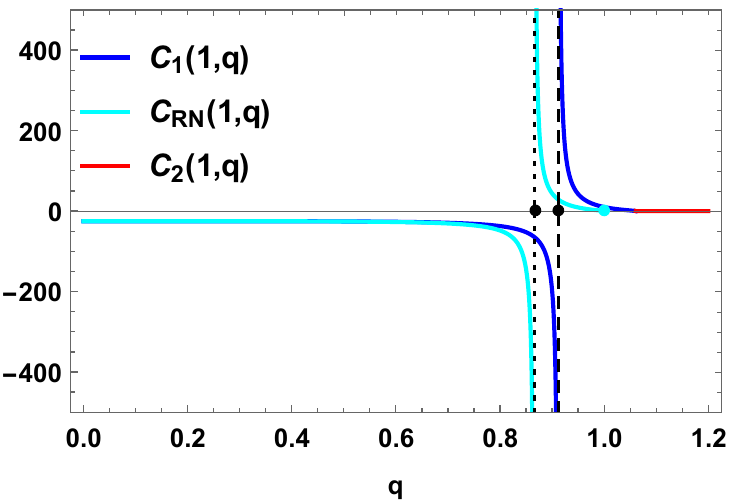}
 \hfill%
    \includegraphics[width=0.4\textwidth]{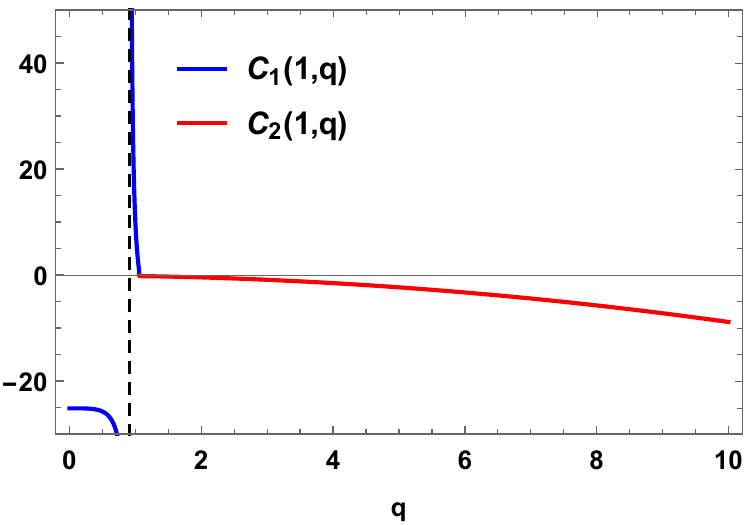}
\caption{(Left) Three heat capacities  $C_1(1,q),~C_2(1,q),~C_{\rm RN}(1,q)$ with $m=1$ as functions of $q\in[0,1.2]$ for the CBH.  There are two Davies points ($\bullet$) and one extremal point (cyan dot) for RNB. (Right) Heat capacity $C_2(1,q)$ with $m=1$ as a function of $q\in[0,10]$ for the NS. }
\end{figure*}

 Furthermore, to test the local thermodynamic stability, one has to compute the heat capacity $C_1=(\partial M/\partial T_1)|_q$.
 It is clear that the thermal  stability (instability) can be achieved when $C_1>0(C_1<0)$.
Two heat capacities  are given by
 \begin{equation}
 C_1(m,q)=-\frac{2\pi r_{+1}^2(m,q)(2r_{+1}^2(m,q)-q^2)}{2r_{+1}^2(m,q)-5q^2},\quad C_{\rm RN}(m,q)=-\frac{2\pi r_{\rm RN}^2(m,q)(r_{\rm RN}^2(m,q)-q^2)}{r_{\rm RN}^2(m,q)-3q^2}.
 \end{equation}
Here, we allow to define the heat capacity $C_2(m,q)$ for the NS by replacing $r_{+1}(m,q)$ by $r_{+2}(m,q)$ in $C_1(m,q)$. As is shown in Fig. 2, the Davies line of $r_{D}(q)=\sqrt{5/2}q$ is derived from imposing the zero numerator of $C_1(1,q)$  and it splits heat capacity into negative and positive ones.
 As is shown in Fig. 5, the CHB heat capacity  with Davies point ($q_D=0.91$) indicates  a similar behavior as the RN heat capacity with  Davies ($q_{D {\rm RN}}=0.87$) and extremal points, whereas the negative  NS heat capacity  has no Davies point and it decreases  as $q$ increases. This implies that the CBH and RNB can be in locally  thermal equilibrium  with  $q$, while the NS cannot be in locally thermal equilibrium  with  $q$.

Finally, the global stability and phase transition are determined by the Helmholtz free energy~\cite{Myung:2007qt,Touati:2022zbm}. If it is positive (negative), it is globally unstable (stable).
Here, we use  the   Helmholtz free energy defined by $F=M-TS$
\begin{equation} \label{free-E}
F_{1/2}(m,q)=\frac{r_{+1/2}(m,q)}{4}+\frac{3q^2}{4r_{+1/2}(m,q)}-\frac{5q^4}{48r_{+1/2}^3(m,q)}\quad F_{\rm RN}= \frac{r_{\rm RN}(m,q)}{4}+\frac{3q^2}{4r_{\rm RN}(m,q)}-q,
\end{equation}
where the latter includes the mass of extremal RNB as a ground state. If one includes the NS singular point as an extremal point, its free energy takes a different form of
\begin{equation}
F_{\rm 1/2NS}(m,q)=F_{1/2}(m,q)-M_{1/2}(m,q_{\rm NS}).
\end{equation}

 \begin{figure*}[t!]
   \centering
  \includegraphics[width=0.4\textwidth]{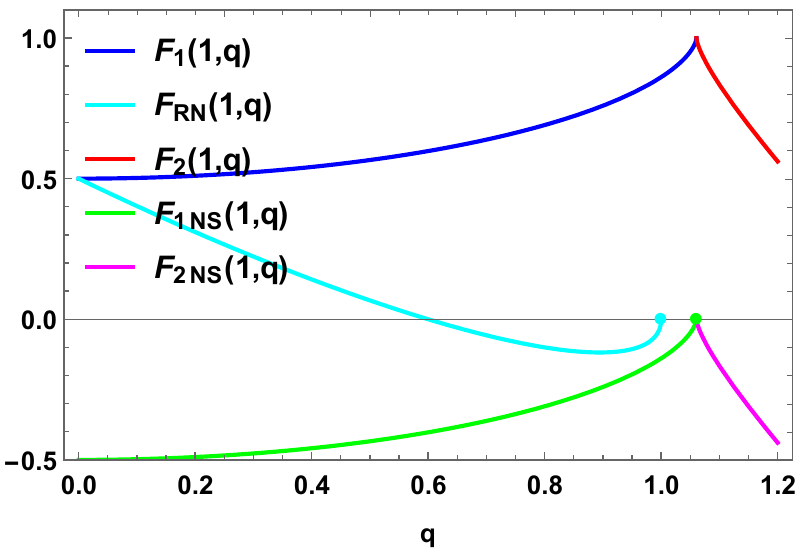}
 \hfill%
    \includegraphics[width=0.4\textwidth]{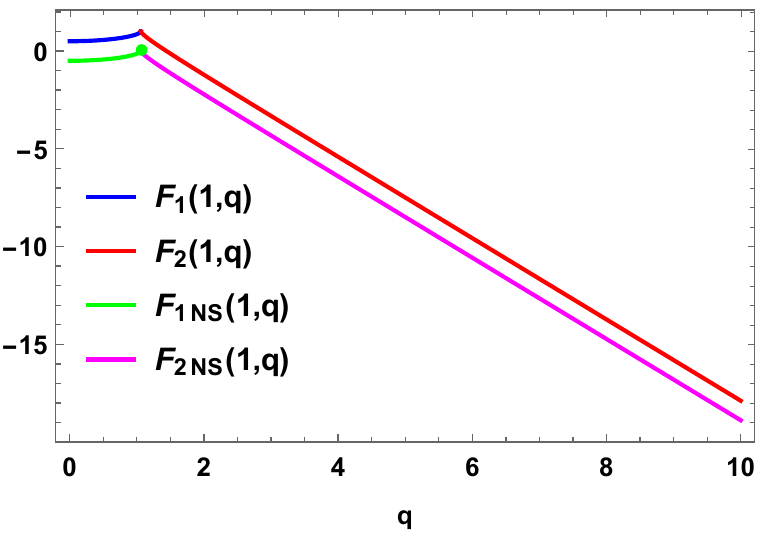}
\caption{(Left) Helmholtz free energy   $F_{1}(1,q),F_{2}(1,q),F_{\rm RN}(1,q)$ as a function of $q\in[0,1.2]$ for the CBH. $F_{\rm 1/2NS}(1,q)$ include the NS singular point as an extremal point.  (Right)
Helmholtz free energy   $F_{2}(1,q)$ for the NS as a monotonically  decreasing function of $q\in[0,10]$.  $F_{\rm 2NS}(1,q)$ includes the NS point. }
\end{figure*}
Observing Fig. 6, one finds that the CHB does not have a globally stable region when excluding the NS point as an extremal point but the NS decreases without limitation as $q$ increases.
For a RNB, its free energy starts with 0.5 and arrives zero at $q=0.6$ and becomes negative. This implies that the RNB can have a globally thermodynamic stability for $0.6<q\le 1$.
If the NS point is included as an extremal point, however,  the CHB has a globally stable region for whole  $q\in[0,1.06]$. In this case, there may be a phase transition from the RN to the CHB because  $S_{\rm RN}(1,q)\le S_{\rm 1NS}(1,q)$ and $ F_{\rm 1NS}(1,q)<F_{\rm RN}(1,q)$.

\section{Discussions}

We have performed extended thermodynamic analysis  of the CHB in $q\in[0,1.06]$ with $m=1$ and the NS in $q\in[1.07,10]$ by introducing two horizons $r_{+1/2}(m,q)$.
We note that  this CHB and NS with scalar hair  has no inner horizon while the RNB has outer and inner horizons.
Let us compare the thermodynamic functions for CHB with those for RNB.

The thermodynamic behavior of the CHB with the NS point is similar to that of RNB in $q\in[0,1]$, where their temperatures  decrease to zero as the charge $q$ increases, their entropies decreases as $q$ increases with terminations,   their heat capacities have a Davies and extremal points, and  their free energies have negative regions.
In this case,  a phase transition   from the RNB to the CHB may occur because  $S_{\rm RN}(1,q)\le S_{\rm 1NS}(1,q)$ and $ F_{\rm 1NS}(1,q)<F_{\rm RN}(1,q)$.
However, there is no overlapping between CHB and RNB (see Table1) because their horizon structures are different (see Fig. 1) and thus,  the only correspondence between CHB and RNB arises from  the $q=0$ limit, leading to the Schwarzschild black hole. This implies that thermodynamic values of   $q=0$ CHB  are   those of $q=0$ RNB, leading to thermodynamic values of Schwarzschild BH.

On the other hand, the CHB without the NS point is quite different from the RNB, implying that the zero temperature cannot be attained  and its free energy is always globally unstable.
Furthermore, the NS thermodynamics may imply that its temperature  has a maximum point at $q=2.37$, its heat capacity remains negative without Davies point, and its negative free energy decreases as $q$ increases without limitation on $q$. However, we are concerned that  the NS inside the NS point has a horizon  to define its thermodynamics.

Finally,  the shadow of BH and NS found from modified gravity theories  were  used to test the EHT results for SgrA$^*$ black hole~\cite{EventHorizonTelescope:2022wkp,EventHorizonTelescope:2022wok,EventHorizonTelescope:2022xqj} and thus, to constrain their parameters~\cite{Vagnozzi:2022moj}.
The CHB was employed to investigating shadow images and photon rings~\cite{Gao:2023mjb}.
We expect that the NS found from the Einstein-Horndeski-Maxwell theory can be used to test the EHT results.
\vspace{1cm}

{\bf Acknowledgments} \\
 \vspace{1cm}

 This work was supported by the National Research Foundation of Korea (NRF) grant
 funded by the Korea government(MSIT) (RS-2022-NR069013).

\end{document}